\documentstyle[aps,floats,graphicx]{revtex}

\begin{document}

\draft


\title{Characteristics of proportional counter filled with
CF$_4$ and Xe additions
}

\author{Ju.\,M.\,Gavriljuk$^1$, A.\,M.\,Gangapshev$^{1}$, V.\,V.\,Kuzminov$^1$
, N.\,Ya.\, Osetrova$^1$,
\\
S.\,I.\,Panasenko$^{2}$, S.\,S.\,Ratkevich$^{2}$}

\address{$^1$ Institute for Nuclear Research, Russian Academy of Sciences,
              Moscow 117312, Russia}

\address{$^2$ Kharkiv National University,
Kharkov 310077, Ukraine}


\maketitle

\begin{abstract}
In this article the measurement results of proportional counter working characteristics
filled with CF$_4$ and Xe ($0\div5$\% Xe) additions at different pressures
($0.8\div14.8$ $at$) are presented. We have found that a bit of Xe addition reduces
working voltage necessary to get the same gas amplification by two times against pure
CF$_4$, it improves as well the resolution of the counter and increases limit gas
amplification by more than ten times.
\end{abstract}

\vskip 0.6 cm


The multicell proportional counter (MCPC) designed to search for weakly-interacting
massive particles (WIMP) was proposed in [1], and is similar to the counter
described in [2]. The suggested experiment belongs to the class of experiments
aimed at the direct detection of WIMP. To the same class of experiments belong recent
experiments with cryogenic, semiconductor and scintillation detectors [3-5].
In such experiments the nuclei recoil energy is detected after interaction of WIMP
with nuclei. The expected energy-release is equal to few keV in our experiment.
It is reasonable to measure such energy-release with help of high-resolution
proportional counter.

One of the possible candidates for dark matter particle is
neutralino -- the lightest stable particle suggested by
supersymmetry. It has spin $s=1/2$ and can interact spin-dependently
with target nuclei with non-zero spin. The intensity of
spin-dependent interaction is more significant for light elements,
such as $^{19}$F ($I=1/2$, $100\%$ isotope abundance) [6]. The proper working
gas with high content of $^{19}$F for MCPC is CF$_4$. CF$_4$ is inflammable
and not toxic. The drift velocity of the electrons in CF$_4$ is high
even under not high electric field [7]. Doe to last property CF$_4$ is
widely used in the detectors intended to detect high flux of
particles in accelerator experiments [8]. Moreover, CF$_4$ can be used
with different gas additions [9,10]. CF$_4$ is a scintillator, which
emits light in the region from ultraviolet (near 160 $nm$) to visible
light, and has about 16\% of Xe scintillation efficiency (s. e.)
[11,12]. Since pure CF$_4$ is transparent for its own ultraviolet
photons, it seems problematical to get high gas amplification in
proportional counters since the photons generated in electron
avalanches stimulate photoeffect on the cathode. The additions of
isobutene or ethane increase a limit of the gas amplification (GA)
by few times [11]. Those results are generally reached for
pressures less than 1 $at$. The characteristics of time-project
chamber filled with pure CF$_4$ at pressure 5 $bar$ are presented in
[12]. The comparison of peak positions of 5.9 keV signal amplitude
distributions which correspond to two different distances between
source and proportional chamber grid allowed to estimate drift
length of the electron to be equal to $9^{+9}_{-3}$ $m$ (electric field
strength 600 $V/cm$). Authors found that the reason for bad resolution
(100\%) caused drift field distortion near $\gamma$-source surface. The
source is placed at a butt-end of the thin (3 $mm$) shift binder. The
majority of photons are absorbed near the source surface, because
the attenuation length of the 5.9 keV photons is $\approx1.5$ $cm$ in CF$_4$ at
pressure of 5 $bar$. Summary resolution of 364 and 388 keV electrons
from conversion electron source $^{113}$Sn equals 25\% (drift length
$\approx5$ $cm$). The resolution of 5.9 keV photon source equals 50\% at
1 $bar$ pressure.
MUNU collaboration has obtained results allowing to build
a detector aimed at the search for neutrino magnetic moment in
antineutrino - electron reactor experiment. The detector is 1m3
time project chamber filled with CF$_4$ at pressure of 3 $bar$. The
transparent chamber vessel is placed into the tank filled with
liquid scintillator used as active shield. In this case CF$_4$ has a
number of advantages against other gases. It has high specific
density (3.68 $g/l$) and high electron density ($1.06\times10^{21}$) at pressure
of 1 $bar$. The absence of high Z elements decreases the multiple
electron recoil scattering as well as distortion of its
trajectories. Moreover the absence of free protons excludes the
background reaction $(\overline\nu  p \to e^ +  n)$.
Photomultipliers scanning liquid scintillator
simultaneously detect scintillations in CF$_4$. The epxploitation
experience of this detector has shown the possibility to use such a
construction as a solar $pp$-neutrino detector with threshold about
100 keV [13]. In our case, the MCPC must have threshold in the
region of about one hundred eV. To have such threshold we need gas
amplification of about 104. The pressure of CF$_4$ must be high enough
to have large mass of the target. The experience of low background
measurements with high-pressure proportional counters [14] shows
that main components of non-ionization background are micro
discharges and current leakage along the isolator surface in high -
voltage circuits. The intensity of this component depends
essentially on the mode and magnitude of the voltage supplied to a
counter (on anode or cathode). In case of MCPC the voltage is
supplied to anodes. The signals are read out from anodes through
high voltage capacitors. Since for this case we have high intensity
of non-ionization background component we need the conditions at
which the voltage is the smallest for a necessary gas
amplification. The latter is possible if we decrease the anode
diameter and find the best addition to CF$_4$.

In this article the proportional counter (PC) characteristics are presented.
The counter was first filled with pure CF$_4$ and later CF$_4$ with Xe addition
at pressures ranging from 0.8 $at$ up to 14.8 $at$.

\begin{center}{\bf MEASUREMENT's CONDITIONS}
\end{center}

The cross section of our PC is shown in fig.1 PC has stainless steel
high-pressure cylindrical body with inner diameter $D=39$ $mm$ and wall thickness of
3 $mm$. Tungsten cathode wires ($D=0.051$ $mm$) are stretched along the side of a
hexahedron parallel to the axis. The high-quality gilded tungsten ($D=0.01$ $mm$)
wire is used as anode. On the whole the inner counter is just one sample of
61 cells at the MCPC. The fiducial length of PC is 17.2 $cm$, total
volume -- 300 $cm^3$, fiducial volume -- 32 $cm^3$.

PC was calibrated with 56.6 $keV$ line of $^{241}$Am source placed on the PC's body.
The positive high voltage is on the anode. The signals are read out off anode through
high voltage separating capacitor by charge sensitive preamplifier (CSP) and sent to
the MCA through the shaping amplifier. The dependence of amplitude of the impulse from
voltage on PC and its time characteristic are measured with oscillograph.
The preparation of PC to the measurement has been carried out with gas-vacuum set-up,
the scheme of which is shown in fig.2. It consists of: empty stainless steel traps
(T$_{8,11,13,14}$), in which liquid gas precipitates at liquid nitrogen temperature
($l.n.t.$); the traps (T$_{12}$) with activated charcoal for gas remains collection
at $l.n.t.$; the flow - type traps with activated charcoal (T$_9$), which is cooled
down to acetone melting temperature (the acetone is cooled by liquid nitrogen).
We divide CF$_4$ and Xe by gas mixture transmission through this trap; pressure gauges
for gas pressure control.

The purification of the gas from electronegative admixture has been realized in
flow - type procedure in the reactor (7) which consist of stainless steel pipe
with absorber [15]. In our work we use CF$_4$, preliminary cleaned from electronegative
admixtures up to $10^{-8}$ oxygen equivalent level [16] and industrial Xe with
residual of $2\times10^{-6}$ of O$_2$. Before beginning our work we pumped out the
set-up and PC down to $10^{-2}$ $torr$.

\begin{center}{\bf RESULTS}
\end{center}

The dependence of gas amplification (GA) of PC filled with pure CF$_4$ on voltage at
pressures 0.8, 1.8, 3.8, 5.8, 8.8, 12.8 $at$ are shown in fig.3. The limitation of
gas amplification for the upper part of the graphic is due to beginning of continuous
discharge process. We take the ion-pair formation energy to be equal to 30 eV for
GA calculation. We read out the spectrum of 59.6 keV line of  $\gamma$-source at
GA$\approx500$ during the same time interval at different pressures. The times of
integration and differentiation are 1.6 $mks$ for all pressures.
Obtained spectra are shown in fig.4. Weak peak of 59.6 keV at 0.8 $at$ may be explained
as a result of wall effect, since the mean range of photoelectron with this energy
equals to 20 $mm$ in CF$_4$. It is necessary to notice that PC's volume is divided by the
grid on two sections with different conditions and electric field intensity.
This could possibly be the reason for the following effects: difference in the time of
charge collection from these regions and difference in amplitudes of the impulse from
events with the same energy release. Both effects are due to  non-transparency of the
grid and different time of interaction with electronegative admixture.
The energy resolution for 59.6 keV line at different pressures is different, and it
changes from 15\% up to 25\%.

Analyzing pulse shape we have found that there are after-impulses in 8 $mks$ after
main 0.5 $mks$-impulse.  The comparative amplitude of after-impulse increases with
voltage, at the same time the after-impulses of next order stay visible. The reason
for this is the photoeffect on the cathode grid and wall from photons generated in
electron avalanche. The comparison of delay time for impulses with calculated drift
time of electrons from the cathode and the wall at 8.8 $at$ ($U=2650$ V) allowed us
to determine that it is generally PC's wall that is the source.

The addition of the hydrocarbon polyatomic gas is usually used to decrease the
voltage on PC and to increase the GA. In our case this addition is  not useful,
because it contains natural tritium (T$_{1/2}=10.8$ $y$, $E=18.7$ keV) causing
additional background. The addition of CO$_2$ (20\%) to CF$_4$ essentially increases
the count characteristic length, increasing, however, at the same the voltage on PC [8].
It is known that small addition of the Xe to pure Ar essentially decreases voltage
necessary to have the same GA as in pure Ar and increases count characteristic
length [18]. The comparison of excitation energy of CF$_4$ molecules with first
ionization potential of Xe allows us to suppose that addition of Xe to CF$_4$ can cause
the same effect.

The dependence of GA on voltage at 8.8 $at$  pressure  of CF$_4$ with addition of Xe
(from 0\% to 5\%)  is shown in fig.5. It is clear that small addition of Xe reduces
the working voltage and increases the limit of GA by more than 10 times
(upper points are limited by the working region of preamplifiers). One can find the
optimal quantity of Xe  analyzing the dependence of voltage on quantity of Xe in CF$_4$
at the same amplitude of impulse in fig.6. It is clear that $1\div2\%$ addition of Xe
is best for our region of pressure. The spectra of 59.6 keV line measured during the
same time interval at pressure 8.8 $at$ for different percentage of Xe are shown
in fig.7. The spectra of 59.6 keV line for mixture with 1\% of Xe at pressures from
1.8 $at$ up to 14.8 $at$ are shown in fig.8.  One can see the total absorption peak
(59.6 keV), escape peak (29.8 keV) on Xe and characteristic lines ($5\div8$ êýÂ) of
Cr, Fe, Ni at al. (elements of the wall and the wire of PC). The energy resolution
of 59.6 keV line equals 13.5\% at 14.8 $at$. Note, that it is necessary to increase
the times of integration and differentiation in amplifier up to 6.4 $mks$ to have good
energy resolution, although electron drift velocity decreases not more than by 2 times
due to decreasing in the voltage. The reason is that in pure CF$_4$ the impulses are
generally formed from single photo- and Compton electrons (point events). In the
mixture CF$_4$+Xe the double-point events appear, in which the second points are caused
by absorption of characteristic radiation emitted by Xe excited due to photo absorption
of primary gamma-photon. In pure CF$_4$ at high pressure the contribution of double-point
events also becomes evident because the probability of secondary interaction of
Compton-photons is high inside PC. The above enumerated effects were confirmed by
increase in energy resolution in pure CF$_4$ for 59.6 keV line from 26.6\% to 8.8\% at
8.8 $at$, when integration and differentiation times change from 1.6 $mks$ to 6.4 $mks$.

\begin{center}{\bf DISSCUSION OF RESULTS}
\end{center}

The increase of GA limit in mixture of CF$_4$ with Xe could be interpreted as evidence
confirming the fact that Xe addition decreases essentially the probability of
photoeffect on PC's wall, caused by ultraviolet photons from CF$_4$.
Those photons are generally induced by (CF$_3^+$)$^*$ è (CF$_4^+$)$^*$ fragments [11]
produced in a gas discharge. (It is possible, that excitation energy of the fragments
is sufficient to ionize Xe atoms, or the presence of the Xe atoms prevent the
formation of  those fragments.)

The discovered mixture may be useful not only in our experiment for dark matter
search with MCPC but for another experiments using CF$_4$ (MUNU-detector, gas detectors
in accelerator experiments).

In conclusion authors would like to thank B.\,M.\,Ovchinnikov and V.\,V.\,Parusov
for their help in getting and clearing CF$_4$ as well as for continuous interest in
our work and productive discussions.

The work is supported by RFBR (grant No 97-02-16051).

\begin{figure}
\begin{center}
\includegraphics[clip,width=3.5 in]{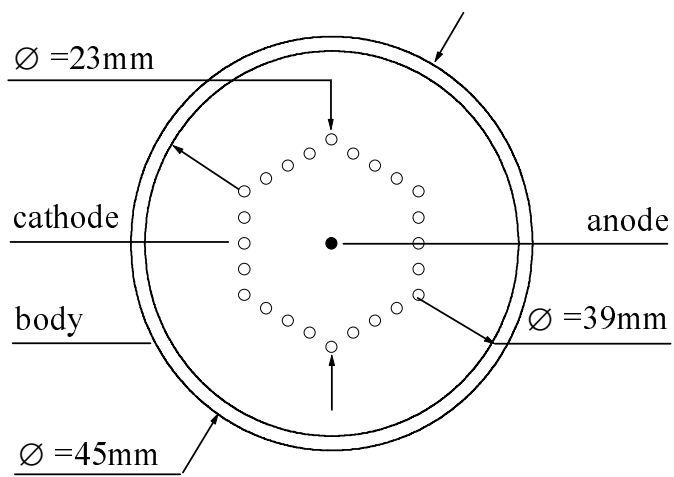}
\vskip 1.6 cm
\caption{The cross section of PC
\label{fig1}}
\end{center}
\end{figure}
%
\begin{figure}
\begin{center}
\includegraphics[clip,width=4.5 in]{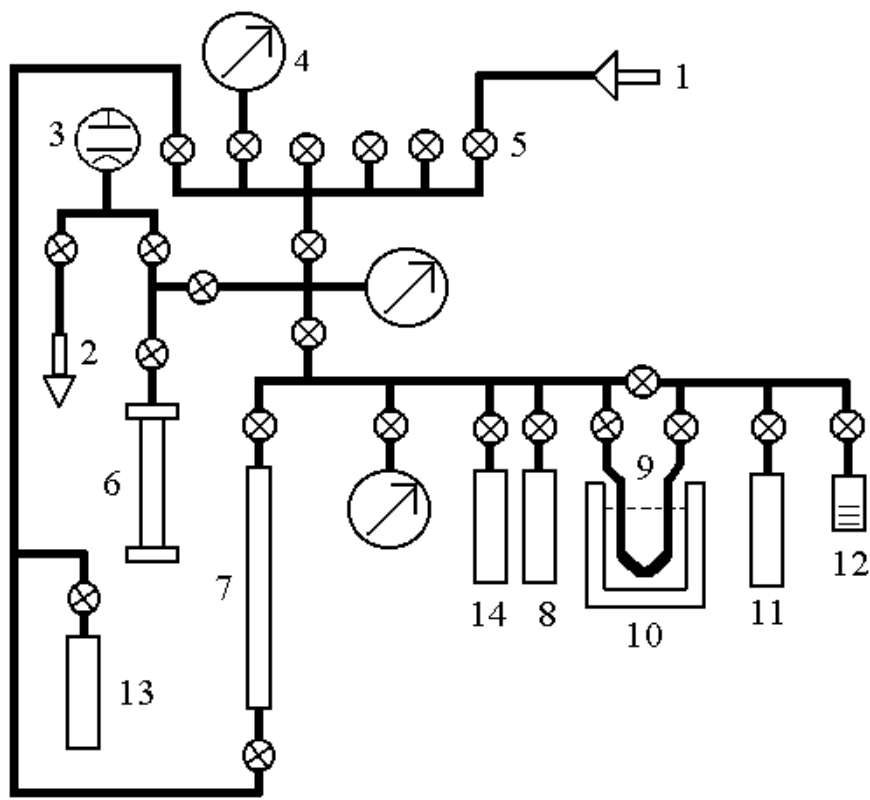}
\vskip 1.6 cm
\caption{The scheme of the installation: 1 - enter for gas;
2 - to the vacuum pump; 3 - thermocouple sensor of vacuum-gauge;
4 - pressure-gauge; 5 -valves; 6 - PC; 7 - reactor with Ni/SiO$_2$ absorber;
8, 11, 13, 14 - traps; 9 - flow - type traps with activated charcoal;
10 - vessel with cooling mixture; 12 - trap with activated charcoal.
\label{fig2}}
\end{center}
\end{figure}
%
\begin{figure}
\begin{center}
\includegraphics[clip,width=4.5 in]{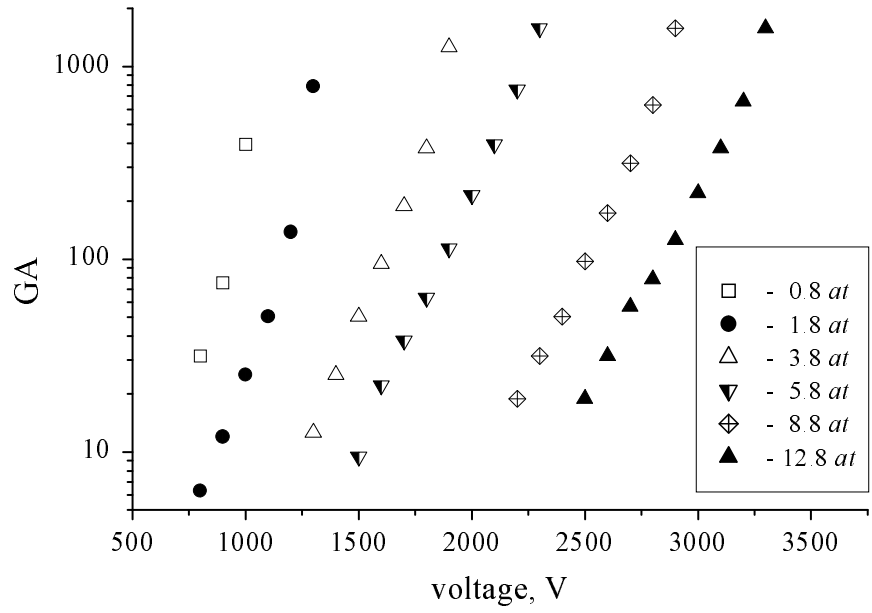}
\vskip 1.6 cm
\caption{Dependents of GA on voltage at different pressures CF$_4$.
\label{fig3}}
\end{center}
\end{figure}
%
\begin{figure}
\begin{center}
\includegraphics[clip,width=4.5 in]{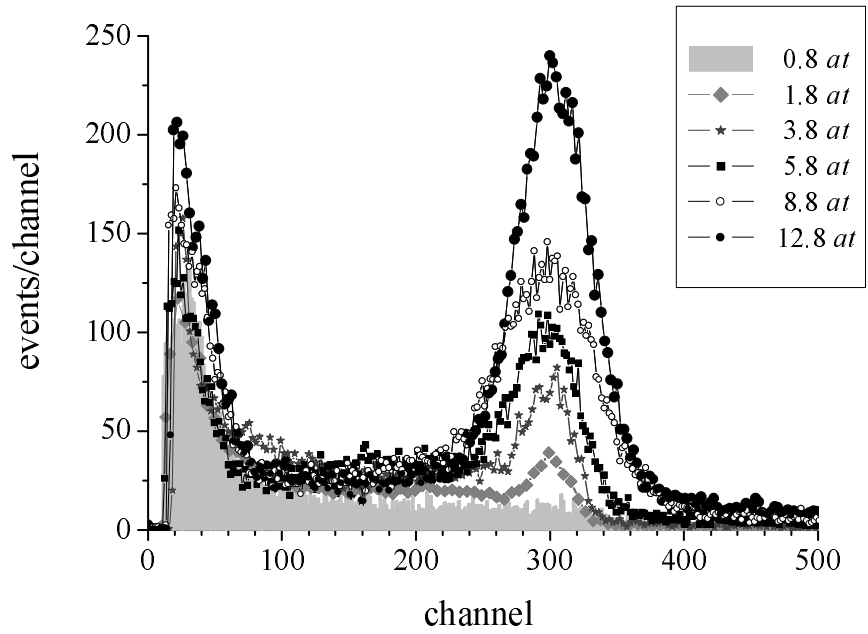}
\vskip 1.6 cm
\caption{Spectra of 59.6 keV line of $\gamma$-source $^{241}$Am at different
pressures of CF$_4$.
\label{fig4}}
\end{center}
\end{figure}
%
\begin{figure}
\begin{center}
\includegraphics[clip,width=4.5 in]{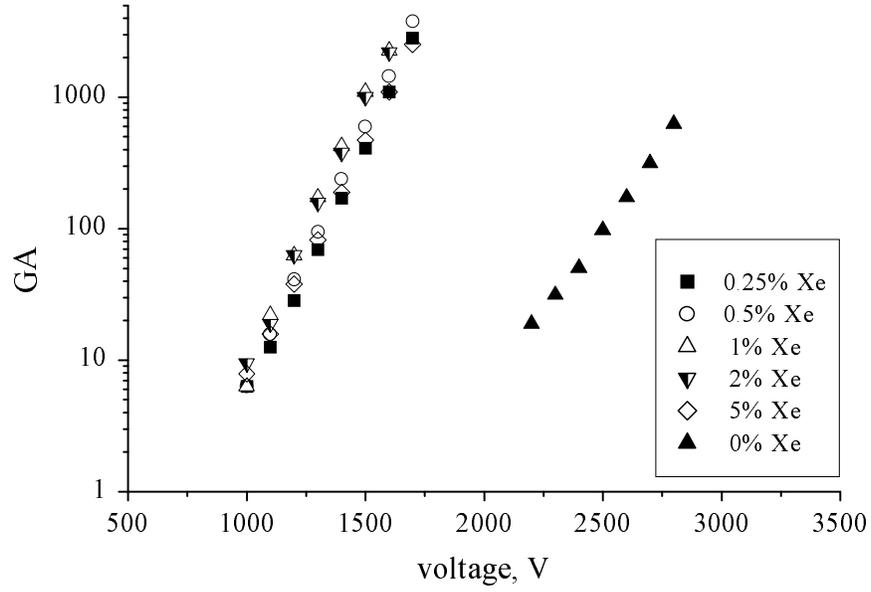}
\vskip 1.6 cm
\caption{Dependence of GA on voltage at 8.8 $at$ pressure of CF$_4$ with addition of
Xe (from 0\% to 5\%).
\label{fig5}}
\end{center}
\end{figure}
%
\begin{figure}
\begin{center}
\includegraphics[clip,width=4.5 in]{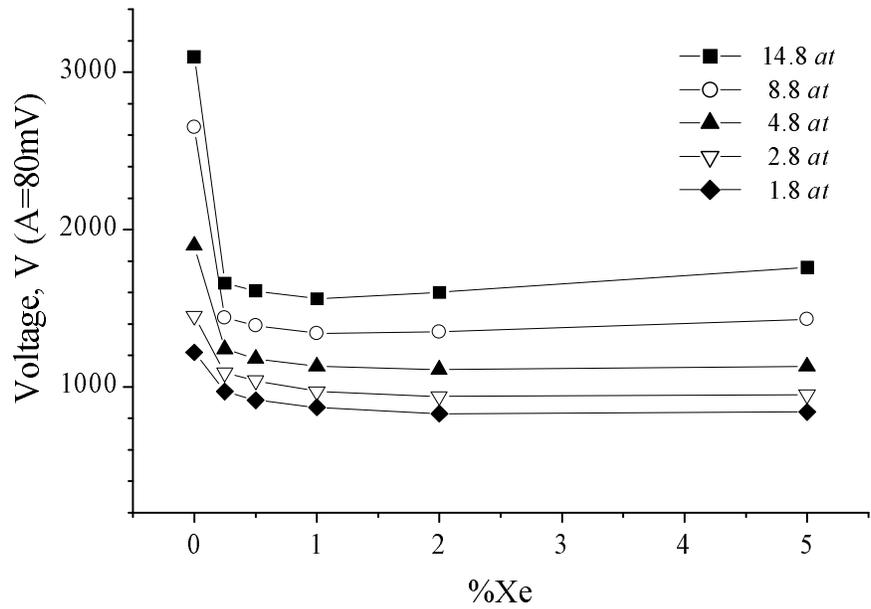}
\vskip 1.6 cm
\caption{Dependence of voltage on quantity of Xe in CF$_4$ at the amplitude of impulses
from 59.9 keV equal to 80 mV.
\label{fig6}}
\end{center}
\end{figure}
%
\begin{figure}
\begin{center}
\includegraphics[clip,width=4.5 in]{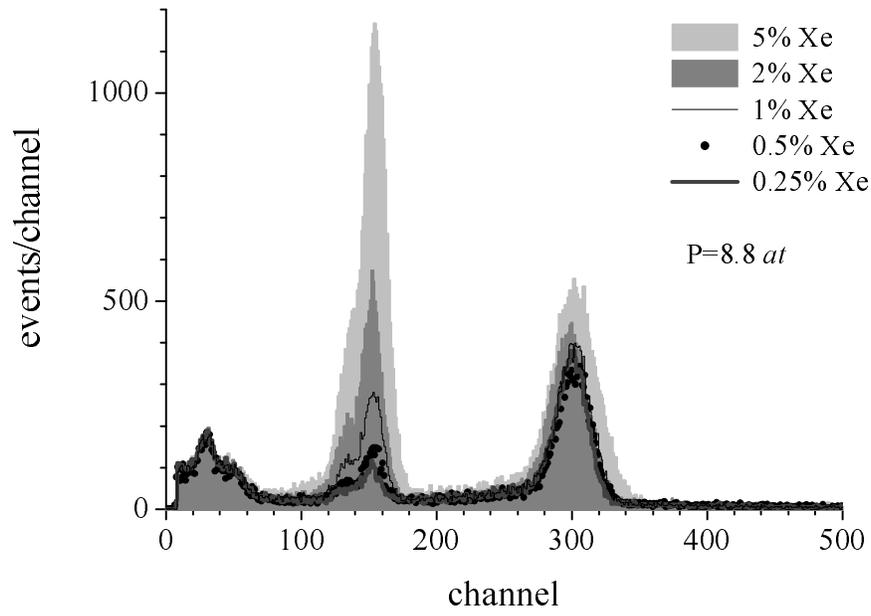}
\vskip 1.6 cm
\caption{Spectra of 59.6 keV line measured during the same time interval
at pressure 8.8 $at$ for different percentage of Xe.
\label{fig7}}
\end{center}
\end{figure}
%
\begin{figure}
\begin{center}
\includegraphics[clip,width=4.5 in]{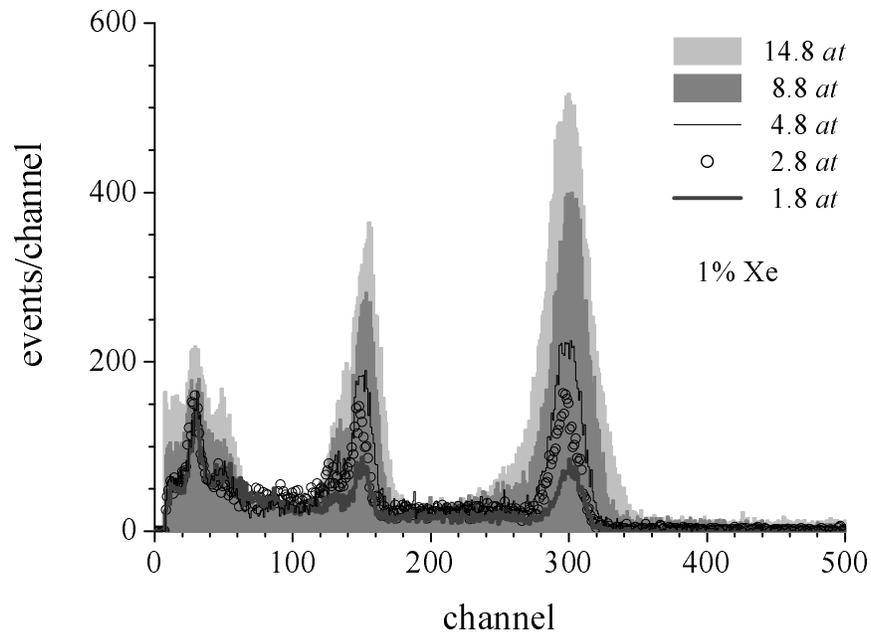}
\vskip 1.6 cm
\caption{Spectra of 59.6 keV line for mixture with 1\% of Xe at pressures
from 1.8 $at$ up to 14.8 $at$.
\label{fig8}}
\end{center}
\end{figure}
\end{document}